\documentclass[onecolumn]{aastex6}
\shorttitle{Improved Lemaitre-Tolman model}
\shortauthors{Del Popolo A. \& Chan M. H.}
\graphicspath{{./}{figures/}}

\usepackage{times}
\usepackage{amsfonts}
\usepackage{amsmath}
\usepackage{amssymb}
\usepackage{natbib}
\usepackage{graphicx}
\usepackage{enumitem}
\usepackage{xcolor}

\def\aap{A\&A}

\def\apjl{ApJL}
\def\apjs{ApJS}

\begin{document}

\title{Improved Lemaitre-Tolman model and the mass and turn-around radius in group of galaxies II: the role of dark energy}

\author{Antonino~\surname{Del Popolo}}
\affiliation{Dipartimento di Fisica e Astronomia, University of Catania, Viale Andrea Doria 6, 95125, Catania, Italy}
\affiliation{Institute of Astronomy, Russian Academy of Sciences, Pyatnitskaya str. 48, 119017 Moscow, Russia}

\author{Man Ho~\surname{Chan}}
\affiliation{Department of Science and Environmental Studies, 
The Education University of Hong Kong, Tai Po, New Territories, Hong Kong
}




\begin{abstract}
In this paper, we extend our previous study \cite{DelPopolo2021} on the Lemaitre-Tolman (LT) model showing how the prediction of the model changes when the equation of state parameter ($w$) of dark energy is modified. In the previous study, it was considered that dark energy was merely constituted by the cosmological constant.  
In this paper, as in the previous study, we also took into account the effect of angular momentum and dynamical friction ($J\eta$ LT model) that modifies the evolution of a perturbation, initially moving with the Hubble flow. 
As a first step, solving the equation of motion, we calculated the relationship between mass, $M$, and the turn-around radius, $R_0$. If one knows the value of the turn-around radius $R_0$, it is possible to obtain the mass of the studied objects. 
%
%

As a second step, we build up, as in the previous paper, a relationship between the velocity, $v$, and radius, $R$. The relation was fitted to data of groups and clusters. Since the relationship $v-R$ depends on the Hubble constant and the mass of the object, we obtained optimized values of the two parameters of the objects studied. 
The mass decreases of a factor of maximum 25\% comparing the $J\eta$ LT results (for which $w=-1$) and the case $w=-1/3$, while the Hubble constant increases going from $w=-1$ to $w=-1/3$. 

%
%

{Finally, the obtained values of the mass, $M$, and $R_0$ of the studied objects can put constraints to the dark energy equation of state parameter, $w$. }
\end{abstract}


 
\section{Introduction}

Several independent observations, and data collected in the last two decades revealed that the universe expansion is accelerated. Observations of supernovae of type IA (SNIa) were the first to indicate the accelerated expansion trend \citep{Perlmutter1998,Riess1998}. This result has been confirmed by several subsequent analyses, like that of the small-scale anisotropies in temperature of the cosmic microwave background radiation \citep{Planck2016}, and other data. The acceleration is interpreted as due to a medium with a negative equation of state (EoS), with EoS parameter $w=P/\rho$, often considered in the range $-1<w<-1/3$. Nevertheless, the nature of dark energy is unknown, observations indicate that it constitutes now about 70\% of matter-energy in the Universe. A plethora of models of dark energy have been introduced, starting from a cosmological constant \citep{Carroll2001}, or a scalar field \citep{Peebles2003}. 

In a previous paper \citep{DelPopolo2021}, we extended  the so called Lemaitre-Tolman model (that we discuss in a while) to take account of the cosmological constant ($w=-1$), angular momentum, and dynamical friction. The model was dubbed $J\eta$LT model. In the present paper, we want to see how the results we discussed in the previous paper, that we will indicate with Paper I \citep{DelPopolo2021} changes for different values of the EoS parameter $w$. 

The mass and the mass-to-light ($M/L$) ratios of group of galaxies are sometimes obtained by means of the Virial theorem. This last is known to give reliable results only in the case the system studied are in dynamical equilibrium. This assumption is often not correct as shown by 
\cite{Niemi2007}, who also showed that $\simeq 20 \%$ of the studied groups were not gravitationally bound.   
Just to give an example, while in the past \cite{Huchra1982}, using the Virial theorem, found values of the mass-to-light ($M/L$) ratios of groups of the order of $\simeq 170 M_{\odot}/L_{B,M_{\odot}}$. More recent measurements, using methods different from the Virial theorem, have given much smaller results (e.g.,  $10-30 M_{\odot}/L_{B,M_{\odot}}$\cite{Karachentsev2004}).

\cite{LyndenBell1981} and \cite{Sandage1986} proposed an alternative approach to the virial theorem based on the Lemaitre-Tolman (LT) model \citep{Lemaitre1933,Tolman1934}.
The quoted method, is used to describe the evolution of a system, similarly to the case of the spherical collapse model (SCM). This model assumes that a system is divided in a series of shells of given radius containing a mass $M$. The spherical perturbation described by the spherical model, initially expands 
following the Hubble flow, until a maximum radius, dubbed turn-around radius, $R_0$, is reached. Later the system starts to collapse. The simplest LT model, taking into account only the gravitational potential energy, is referred to as SLT (standard LT model) to distinguish it from extension of the model, is characterized by a central region in equilibrium, surrounded by the region that expands to a maximum radius and then recollapses.
A group of galaxies dominated by a central one or a binary system
is well described by the LT model. 

In the simplest case, in which one considers only the effect of the gravitational potential (SLT model), as shown by \cite{Sandage1986}, \cite{Peirani2006,Peirani2008} the mass is given by
\begin{equation}
\label{eq:LT}
M=\frac{\pi^2 R_0^3}{8GT_0^2}= 3.06 \times 10^{12} h^2 R_0^3 M_{\odot},
\end{equation}
where $T_0$ is the age of the universe, and $h$ is the Hubble constant in units of 100 $\rm km s^{-1}/Mpc$. This simple model was applied to the Local Group \citep{Sandage1986} and to the Virgo cluster \citep{Hoffman1980, Tully1984,Teerikorpi1992}. 
The first extension of the model, taking into account the effect of the cosmological constant was presented by \cite{Peirani2006,Peirani2008}. The model is dubbed in the following "modified LT model" (MLT). They applied the model to Virgo cluster, the pair M31-MW, M81, the Centaurus A-M83 group, the IC342/Maffei-I group, and the NGC 253 group. 
This model shows a different relation between the mass $M$, and $R_0$, namely 
\begin{equation}
\label{eq:LTt}
M=\frac{1.69945 R_0^3}{GT_0^2}= 4.22 \times 10^{12} h^2 R_0^3 M_{\odot}
\end{equation}
%
%
\cite{Peirani2006,Peirani2008}, proposed also another method to calculate the mass. They built up a velocity-distance relationship, $v-R$, describing the kinematic status of the systems studied. If one can determine the velocities $v$, and distance $R$ for the members of the groups studied, one can obtain the mass, $M$, and the Hubble constant by means of a non-linear fit of the $v-R$ relation to the data. \cite{DelPopolo2021} extended the work of \cite{Peirani2006,Peirani2008} by taking into account the effect of angular momentum (JLT model) and dynamical friction (J$\eta$LT model)\footnote{The effect of the specific angular momentum $J$, and dynamical friction on the turn-around, the threshold of collapse, the clusters of galaxies structure and evolution, their mass function, and their mass-temperature relation, have been studied in several papers \citep{DelPopolo1998,DelPopolo1999,DelPopolo2000,DelPopolo2006,DelPopolo2006a,DelPopolo2006b,DelPopolo2017,DelPopolo2019,DelPopolo2020}. 
}.

In the present paper, we are interested in how the change of the EoS parameter $w$ influences the result of the $J\eta$ LT model. 

As a first step, we will solve the equations of motion of the system to see how the $M-R_0$ relation changes with $w$. Then, we will build up the $v-R$ relation, as done in \cite{Peirani2006,Peirani2008}, \cite{DelPopolo2021}, and fit it to the data of the Virgo cluster, the pair M31-MW, M81, the Centaurus A-M83 group, the IC342/Maffei-I group, and the NGC 253 group.

{ 
The paper is organized as follows. Section~\ref{sec:Model} introduces the model, and shows how to solve it. In Section~\ref{sec:VR_relation}, we find the velocity-radius relations for the case $w=-2/3$ ($J\eta \rm LT_{-2/3}$ model), and the case $w=-1/3$ ($J\eta \rm LT_{-1/3}$ model).
In Section~\ref{sec:ApplicationNearGroups}, the $v-R$ relation was applied to groups and clusters of galaxies.
In Section~\ref{sec:DMconstraints}, we put constraints on the dark energy equation of state parameter. Section~\ref{sec:Conclusions} is devoted to conclusions.
}

\section{Model}
\label{sec:Model}

As previously reported, the SCM introduced by \cite{Gunn1972} is a simple method to study analytically the non-linear evolution of perturbations of dark matter (DM) and dark energy (DE).
It describes the evolution of a spherically symmetric perturbation which is initially following the Hubble flow, and then detaches from it and collapses forming a structure. In the \cite{Gunn1972}
model was taken into account only the gravitational potential of the mass, and matter moves in a radial fashion 
\citep{Gunn1972,Gunn1977}.
SCM was improved in several papers adding the effect of the cosmological constant \citep{Lahav1991}, tidal angular momentum \citep{Peebles1969,White1984}, 
random angular momentum
\citep{Ryden1987,Gurevich1988a,Gurevich1988b,White1992,Sikivie1997,Nusser2001,Hiotelis2002,  
LeDelliou2003,Ascasibar2004,Williams2004,Zukin2010}, and dynamical friction  \citep{AntonuccioDelogu1994,Delpopolo2009}. The model has been used to study a series of issues. 
\cite{DelPopolo2013a,DelPopolo2013b} studied the effects of shear and rotation for smooth DE models, while \cite{Pace2014b} studied them in clustering DE cosmologies, and 
\cite{DelPopolo2013c} in Chaplygin cosmologies. 
Several authors  \citep{Bernardeau1994,Bardeen1986,Ohta2003,Ohta2004,Basilakos2009,Pace2010,Basilakos2010} studied the SCM with negligible DE perturbations, and others  \citep{Mota2004,Nunes2006,Abramo2007,Abramo2008,Abramo2009a,Abramo2009b,
Creminelli2010,Basse2011,Batista2013} took account of DE fluid perturbation. The main parameters of the SCM are mass independent, but if the role of shear and rotation are taken into account they become mass dependent 
\citep{Pace2010,DelPopolo2013b}.


The equation of motion of the system are given following \cite{Peebles1993}, \cite{Bartlett1993}, \cite{Lahav1991}, \cite{DelPopolo1998}, \cite{DelPopolo1999}, \cite{DelPopolo2006b} by the equation given in \cite{DelPopolo2021}:
\begin{equation}
\label{eq:coll1}
 \frac{{\rm d}v_R}{{\rm d}t} = -\frac{GM}{R^2} + \frac{J^2}{R^3}-\frac{1+3w}{2} \Omega_{\Lambda} H_0^2 (\frac{a_0}{a})^{3(1+w)} R
-\eta\frac{{\rm d}R}{{\rm d}t}\,,
\end{equation}
where $J=\frac{L}{M}$ is the specific angular momentum, $L$ is the angular momentum and takes into account ordered angular momentum generated by tidal torques and random angular momentum (see Appendix C.2 of \cite{Delpopolo2009}), $\Omega_{\Lambda}=\frac{\rho_{\Lambda}}{\rho_c}$, where $\rho_c$ is the critical density, $\rho_{\Lambda}$ is the density related to the cosmological constant $\Lambda$, $\eta$ (explicitly given in \cite{Delpopolo2009} (Appendix D, Eq. D5)) the dynamical friction coefficient per unit mass, where $w$ is the DE equation of state (EoS) parameter. DE is modeled by a fluid with an EoS given by $P=w \rho$, where $\rho$ is the energy density. $a$ is the expansion parameter. Eq.\eqref{eq:coll1} satisfies the equation
\begin{equation}
\label{eq:hubbo}
H=\frac{\dot a}{a}=H_0 \sqrt{\Omega_{\rm m} (\frac{a_0}{a})^3+\Omega_{\Lambda}\left(\frac{a_0}{a}\right)^{3(1+w)} }.
\end{equation}

Assuming that $J=k R^{\alpha}$, with $\alpha=1$, in agreement with \cite{Bullock2001}\footnote{In that paper $\alpha=1.1 \pm 0.3$}, and $k$ constant, in terms of the variables $y=R/R_0$, $t=x/H_0$, Eq.\eqref{eq:coll1} 
can be written as
\begin{equation}
\label{eq:princ}
\frac{d^2y}{dx^2}=-\frac{A}{2y^2}-\frac{1+3w}{2}\Omega_{\Lambda} y (\frac{a_0}{a})^{3(1+w)}+\frac{K_j}{y}-\frac{\eta}{H_0}\frac{dy}{dx},
\end{equation}
where $K_j=k\frac{1}{(H_0R_0)^2}$, $A=\frac{2GM}{H_0^2 R_0^3}$.
%
In terms of $\xi=a/a_0$, and $x$, Eq. (\ref{eq:hubbo}) can be written as 
\begin{equation}
\label{eq:hubb2}
\frac{d \xi}{dx}= \sqrt{\frac{\Omega_m}{\xi}+\frac{\Omega_{\Lambda}}{\xi^{1+3w}}}.
\end{equation}
or separating the variables, as
\begin{equation}
\label{eq:hubb3}
x= \int_0^1 \frac{d \xi}{\sqrt{\frac{\Omega_m}{\xi}+\frac{\Omega_{\Lambda}}{\xi^{1+3w}}}}.
\end{equation}

Eq.\eqref{eq:princ} has a first integral, given by
%
%

\begin{align} 
\label{eq:princ1}
u^2&=\left(\frac{dy}{dx}\right)^2
=\frac{A}{y}-\frac{1+3w}{2}\Omega_{\Lambda} y^2 (\frac{a_0}{a})^{3(1+w)} \nonumber\\
&+2K_j \log{y} 
-2\frac{\eta}{H_0} \int \left(\frac{dy}{dx}\right)^2 dx+K
\end{align}

where $K=\frac{2E}{(H_0R_0)^2}$, and $E$ is the energy per unit mass of a shell.

Eq.\eqref{eq:princ}, and Eq.\eqref{eq:hubb2} were solved as described in \cite{Peirani2006, Peirani2008}. This can be done in a couple of ways. One way is to get the value of 
scale parameter and the corresponding time for a given redshift, from Eq. (\ref{eq:hubb2}). 
Moreover, the gravitational term dominates at high redshift. By means of a Taylor expansion  
the initial conditions can be obtained. The parameter $A$, which is the goal of the calculation,  is varied until the condition $\frac{dy}{dx}=0$, and $y=1$ are satisfied. 

The second way to get $A$, is based on the use of the equation for the velocity (Eq.\eqref{eq:princ1}), as described in \cite{DelPopolo2021}. 

In \cite{DelPopolo2021}, we indicated the general solution of Eq. (\ref{eq:princ}), the one with
$w=-1$, $\Omega_{\Lambda}=0.7$, $\Omega_{\rm m}=0.3$, $K_j=0.78$, $\eta/H_0=0.5$, with $J\eta$LT.
As shown in \cite{DelPopolo2021}, we got $A=6.05$, or $M=\frac{2.8111 R_0^3}{GT_0^2}=6.9843 \times 10^{12} h^2 R_0^3 M_{\odot}$.


We solved Eq.\eqref{eq:princ}, and Eq.\eqref{eq:hubb2}, also for the case $w=-2/3$ (that we dubb $J\eta$LT$_{-2/3}$), and $w=-1/3$ (that we dubb $J\eta$LT$_{-1/3}$). In the case $J\eta$LT$_{-2/3}$, $A=5.83$, corresponding to $M=6.73 \times 10^{12} h^2 R_0^3 M_{\odot}$, and in the case $J\eta$LT$_{-1/3}$ , $A=5.32$, corresponding to $M=6.14 \times 10^{12} h^2 R_0^3 M_{\odot}$.  

\begin{figure}[t]  
	\centering
	\includegraphics[scale=0.30]{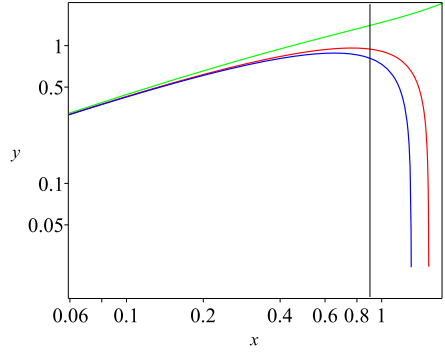}
	\caption{
		Evolution of shells with different energies ($K$) in the $J\eta \rm LT_{-2/3}$ model. The green, red, and blue lines correspond to $K=-4.8$, $K=-6.18$, and $K=-6.5$, respectively. The vertical black line is the time $x=0.903$, at which $R=R_0$.
	}
	\label{fig:RadiusEvoluion_diff_K}
\end{figure}

In Fig.\ref{fig:RadiusEvoluion_diff_K}, we plotted three solutions of the case 
$J\eta \rm LT_{-2/3}$ for different values of $K$. The red line represents the case $K=-6.18$, and represents the solution that just reached turn-around. The blue line has $K=-6.5$, and reached turn-around in the past. The value of $K=-5.327$ is a threshold value. For $K<-5.327$, turn-around and collapse are allowed, while are not allowed for $K>-5.327$, as the case of the green line, having $K=-4.8$.
The vertical black line is the time $x=0.903$, at which $R=R_0$.

\begin{figure*}[t]
	\centering
   \includegraphics[scale=0.25]{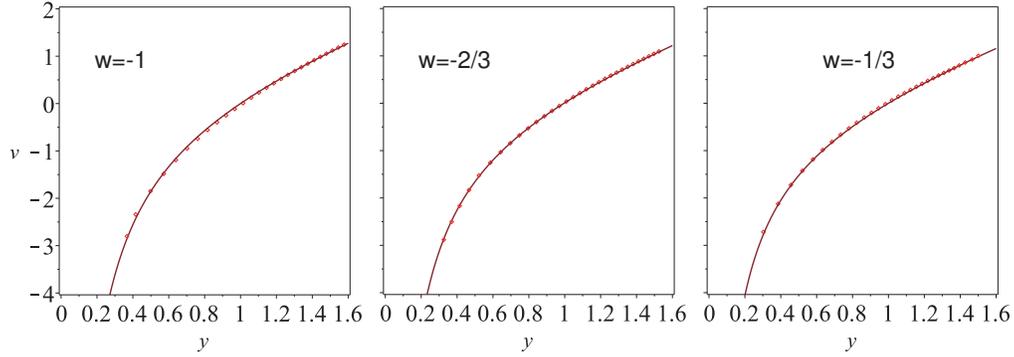}
	\caption{
		Velocity profile for different $w$. The left panel shows the $J\eta$LT model ($w=-1$). The central panel, the $J\eta$LT$_{-2/3}$ case, and the right panel, the $J\eta$LT$_{-1/3}$ case.
The data points are solutions of Eq. (\ref{eq:princ}) for different values of $K$ (see Eq. (\ref{eq:princ1})). In Paper I, and its Fig. 2, and in the text of the present paper, there is a deeper discussion. The lines are fit to the data, in the form $v=-b/y^n+by$, whose parameters can be found in Table \ref{tab:fit_param}. 		
	}
	\label{fig:agnularMom_dynFric}
\end{figure*}

\section{Determination of the velocity-radius relation}
\label{sec:VR_relation}

As we already reported, \cite{Peirani2006,Peirani2008}, and \cite{DelPopolo2021} found a relation between the velocity and the radius, $v-R$, that was fitted to the data of some groups of galaxies. The relation depends from the Hubble constant, $H$, and the mass, $M$, of the system. After $H$, and $M$ are obtained it is also possible to obtain the turn-around radius $R_0$, by means of the relation, already written, $A=\frac{2GM}{H_0^2 R_0^3}$. As discussed, the coefficient $A$ is obtained solving the equations of motions.  
We already discussed how to obtain the $v-R$ relation in \cite{DelPopolo2021}, and it was also discussed in \cite{Peirani2006,Peirani2008}. For reader's convenience, we recall how we get the quoted relation. Let's consider Fig.\ref{fig:RadiusEvoluion_diff_K}. The plot represents three solutions that are obtained from Eq. (\ref{eq:princ}), and Eq. (\ref{eq:hubb2}), for three different values of $K$. The black vertical line in Fig.\ref{fig:RadiusEvoluion_diff_K} corresponds to $x=0.903$, as can be obtained from Eq. (\ref{eq:hubb3}). The intersection of the black line with each curve $y(x)$, corresponding to a given $K$, gives a succession $y_K (0.903)$.   
Moreover, solving Eq. (\ref{eq:princ}), and Eq. (\ref{eq:hubb2}), one can obtain the velocity $v(x)$, and again a succession $v_k(0.903)$.   
We will get a couple of value $(y,v)$ for each intersection of the vertical line with the curves, and considering all the intersections, we have   $y_k(0.903),v_k(0.903)$. In Fig.\ref{fig:agnularMom_dynFric}, we plotted in red the succession of points in the cases $J\eta$LT (left panel),
$J\eta \rm LT_{-2/3}$ (central panel), and $J\eta \rm LT_{-1/3}$ (right panel). 
The series of points can be fitted with a relation of the form $v=-b/y^n+by$. In Fig.\ref{fig:agnularMom_dynFric}, the fit is represented by the brown lines. The parameters of the fit, $b$, and $n$ are reported in Table \ref{tab:fit_param}, for the three cases considered ($J\eta$LT ($w=-1$), $J\eta \rm LT_{-2/3}$, and $J\eta \rm LT_{-1/3}$). 
The fit $v=-b/y^n+by$ can be written in terms of the physical units as
\begin{equation} 
\label{eq:recov}
v(R)=-b H_0 R_0 \left(\frac{R_0}{R}\right)^n+b H_0 R.
\end{equation}
Recalling that $R_0=(\frac{2GM}{H_0^2})^\frac{1}{3}$, substituting in the previous equation, we get
\begin{equation}
v(R)=-b\frac{H_0}{R^n} \left(\frac{2GM}{A H_0^2}\right)^\frac{n+1}{3}+bH_0R .
\end{equation}
Using the parameters in Table \ref{tab:fit_param}, in the case $J\eta \rm LT_{-2/3}$, we have
\begin{equation}
\label{eq:Om}
v(R)=-\frac{-0.69 H_0}{R^{0.82}} \left(\frac{GM}{H_0^2}\right)^{0.61}+1.32 H_0 R ,
\end{equation}
In the case of the $J\eta \rm LT_{-1/3}$ case, we have:
\begin{equation}
\label{eq:Omi}
v(R)=-\frac{-0.74 H_0}{R^{0.75}} \left(\frac{GM}{H_0^2}\right)^{0.58}+1.29 H_0 R ,
\end{equation}
and in the case $J\eta$LT ($w=-1$), which was already studied in Paper I, we have
\begin{equation}
\label{eq:eta}
v(R)=-\frac{-0.59 H_0}{R^{0.91}} \left(\frac{GM}{H_0^2}\right)^{0.64}+1.34 H_0 R .
\end{equation}
Notice that this last $v-R$ relation (Eq. (\ref{eq:eta})) is slightly different from that of Paper I, where the first term was 0.66, and here 0.59. We checked again all the fits and noticed that in the case $w=-1$ there was a small discrepancy.
This produces also a change of the mass values and $R_0$ in Table (\ref{tab:hRM_fits}) for the case $J\eta$LT.

As already reported, in Fig.\ref{fig:agnularMom_dynFric}, we plot, from left to right, the velocity profiles of the cases $J\eta$LT (left panel), $J\eta \rm LT_{-2/3}$ (central panel), and $J\eta \rm LT_{-1/3}$, using adimensional variables.

All the previous equations satisfy the condition $v(R_0)=0$. In the following, we will apply Eq.\eqref{eq:Om}, Eq.\eqref{eq:Omi} to some groups of galaxies and clusters. 
The parameters of the different models that were described in this paper, plus the case $J\eta$LT studied in paper I are summarized in Table \ref{tab:fit_param}. The first line corresponds to the $J\eta$LT model. The second line to the $J\eta \rm LT_{-2/3}$ model, and, the last line to the $J\eta \rm LT_{-1/3}$ case.
\begin{table}[ht]
	\begin{center}
		\begin{tabular}{l|ccccc}
			\hline
			model
			&$\eta/H_0$
			&\multicolumn{1}{c}{$K_{J}$}
			&\multicolumn{1}{c}{$b$}	
			&\multicolumn{1}{c}{$n$}
			&\multicolumn{1}{c}{$A$}\\	
			\hline
	        \multicolumn{1}{c}{J$\eta$LT } 
			&\multicolumn{1}{|c}{$0.5$} 
			&\multicolumn{1}{l}{$0.78$} 
			&\multicolumn{1}{l}{$1.34$}
			&\multicolumn{1}{l}{$0.91$} 
			&\multicolumn{1}{l}{$6.05$}\\															
            \multicolumn{1}{c}{$J\eta$LT$_{-2/3}$} 
			&\multicolumn{1}{|c}{0.5} 
			&\multicolumn{1}{l}{$0.78$} 
			&\multicolumn{1}{l}{$1.32$}
			&\multicolumn{1}{l}{$0.82$} 
			&\multicolumn{1}{l}{$5.83$}\\	
			\multicolumn{1}{c}{$J\eta$LT$_{-1/3}$} 
			&\multicolumn{1}{|c}{0.5} 
			&\multicolumn{1}{l}{$0.78$}  
			&\multicolumn{1}{l}{$1.29$}
			&\multicolumn{1}{l}{$0.75$} 
			&\multicolumn{1}{l}{$5.32$}\\

			\hline								
		\end{tabular}
	\end{center}
	\caption{
	The constant $A$, and the fitting parameters $b$, and $n$ of the velocity-distance ($v-R$) relations, for the $J\eta$LT, the $J\eta \rm LT_{-2/3}$, and $J\eta \rm LT_{-1/3}$
models.
}
\label{tab:fit_param}	
\end{table} 

Table \ref{tab:fit_param}, as well as Fig.\ref{fig:agnularMom_dynFric} shows that with decreasing $w$, from $w=-1$ of the case $J\eta$LT to $w=-2/3$ of the case $J\eta \rm LT_{-2/3}$, and $w=-1/3$ of the case $J\eta \rm LT_{-1/3}$ the values of the parameter $A=\frac{2GM}{H_0^2 R_0^3}$ decreases, and the velocity profile flattens. For a fixed value of $R_0$, and $H_0$, the mass of the structures decreases.  
\begin{table*}[t]
  \centering
  \caption{
  {Values of the parameters of the examined groups. The rows 1-3 represent the value of the Hubble constant in units of 100 $\rm km s^{-1}/Mpc$ for the cases $J\eta \rm LT_{-2/3}$, $J\eta \rm LT_{-1/3}$, and $J\eta$LT. The masses in units of $10^{12} M_{\odot}$, and for the same cases, are presented in rows 4-6, while in the rows 7-9 the values of the turn-around radius, $R_0$, in Mpc.  The rows 10-12 give the velocity dispersion obtained as the other parameters by fitting the data to the $v-R$ relation.}
}
\begin{tabular}{l|cccccc}
\hline
\scriptsize{}
&\scriptsize{\bf{M31/MW}}
&\scriptsize{\bf{M81}}
&\scriptsize{\bf{NGC 253}} 
&\scriptsize{\bf{IC 342}}
&\scriptsize{\bf{CenA/M83}}
&\scriptsize{\bf{Virgo}}\\
\hline   
\scriptsize{h ($J\eta$LT$_{-1/3}$)} 
&\scriptsize{$0.75 \pm 0.04$}  &\scriptsize{$0.7 \pm 0.04$}  &\scriptsize{$0.67 \pm 0.06$}&\scriptsize{$0.60 \pm 0.10$}  &\scriptsize{$0.59 \pm 0.04$}  &\scriptsize{$0.69 \pm 0.09$} \\  
\scriptsize{h ($J\eta$LT$_{-2/3}$)}
&\scriptsize{$0.72 \pm 0.04$}  &\scriptsize{$0.67 \pm 0.04$}  &\scriptsize{$0.65 \pm 0.06$} &\scriptsize{$0.57 \pm 0.10$}  &\scriptsize{$0.57 \pm 0.04$} &\scriptsize{$0.64 \pm 0.08$} \\
\scriptsize{h ($J\eta$LT)} 
&\scriptsize{$0.69 \pm 0.04$}  &\scriptsize{$0.65 \pm 0.04$} &\scriptsize{$0.63 \pm 0.05$}  &\scriptsize{$0.55 \pm 0.10$}  &\scriptsize{$0.55 \pm 0.04$}&\scriptsize{$0.59 \pm 0.09$} \\ 
\scriptsize{M($J\eta$LT$_{-1/3}$) [$10^{12} M_{\odot}$]}
&\scriptsize{$3.27 \pm 0.50$} &\scriptsize{$1.40 \pm 0.10$}  &\scriptsize{$0.21 \pm 0.10$} &\scriptsize{$0.28 \pm 0.10$}&\scriptsize{$2.81 \pm 0.50$}&\scriptsize{$1681 \pm 200$} \\
\scriptsize{M($J\eta$LT$_{-2/3}$) [$10^{12} M_{\odot}$]}
&\scriptsize{$3.72 \pm 0.50$}  &\scriptsize{$1.53 \pm 0.10$}  &\scriptsize{$0.24 \pm 0.15$}&\scriptsize{$0.312 \pm 0.12$} &\scriptsize{$3.17 \pm 0.50$} &\scriptsize{$1760 \pm 200$} \\ 
\scriptsize{M($J\eta$LT) [$10^{12} M_{\odot}$]} 
&\scriptsize{$4.31 \pm 0.40$}  &\scriptsize{$1.68 \pm 0.10$} &\scriptsize{$0.31 \pm 0.10$} &\scriptsize{$0.35 \pm 0.10$}&\scriptsize{$3.65 \pm 0.40$}&\scriptsize{$1844 \pm 200$} \\ 
\scriptsize{R($J\eta$LT$_{-1/3}$) [Mpc]}
&\scriptsize{$0.99 \pm 0.10$}  &\scriptsize{$0.78 \pm 0.05$} &\scriptsize{$0.43 \pm 0.10$} &\scriptsize{$0.51 \pm 0.09$} &\scriptsize{$1.10 \pm 0.08$}&\scriptsize{$8.35 \pm 0.80$}\\
\scriptsize{R($J\eta$LT$_{-2/3}$) [Mpc]}
&\scriptsize{$1.02 \pm 0.10$} &\scriptsize{$0.80 \pm 0.05$} &\scriptsize{$0.44 \pm 0.10$} &\scriptsize{$0.52 \pm 0.09$}&\scriptsize{$1.14 \pm 0.08$} &\scriptsize{$8.62 \pm 0.80$}\\
\scriptsize{R($J\eta$LT) [Mpc]}
&\scriptsize{$1.08 \pm 0.10$} &\scriptsize{$0.83 \pm 0.05$} &\scriptsize{$0.48 \pm 0.10$} &\scriptsize{$0.55 \pm 0.09$}&\scriptsize{$1.20 \pm 0.08$} &\scriptsize{$9.12 \pm 0.80$}\\
\scriptsize{$\sigma(J\eta \rm LT_{-1/3})$ [km/s]}
&\scriptsize{$38.29$}  &\scriptsize{$53.19$} &\scriptsize{$44.75$} &\scriptsize{$33.41$} &\scriptsize{$45.96$}&\scriptsize{$352.84$}\\
\scriptsize{$\sigma(J\eta \rm LT_{-2/3})$ [km/s]}
&\scriptsize{$38.52$} &\scriptsize{$53.90$} &\scriptsize{$44.17$} &\scriptsize{$34.17$}&\scriptsize{$45.91$} &\scriptsize{$353.69$}\\
\scriptsize{$\sigma(J\eta LT)$ [km/s]}
&\scriptsize{$38.8$} &\scriptsize{$54.77$} &\scriptsize{$45.9$} &\scriptsize{$34.48$}&\scriptsize{$44.81$} &\scriptsize{$355.1$}\\
\hline    
    \end{tabular}%
\label{tab:hRM_fits}%
\end{table*}%

\section{Application to data from near groups and clusters of galaxies}
\label{sec:ApplicationNearGroups}

As we already discussed, we obtained a $v-R$ relation for the case $J\eta \rm LT_{-2/3}$, (Eq. \ref{eq:Om}), $J\eta \rm LT_{-1/3}$ (Eq. (\ref{eq:Omi})), and in paper I that for the case $J\eta$LT (Eq. (\ref{eq:eta}))\footnote{For precision's sake, in the present paper we recalculated also the $v-R$ relation, and performed the fit to the quoted groups for the $J\eta$LT case}. These relations depend on mass, $M$, and the Hubble constant, $H$. We will now fit the quoted relations to near groups, and to the Virgo cluster. To perform the fit, we need for each galaxy its distance and velocity with respect to the center of mass of the system studied. The data were obtained, as described in \cite{Peirani2006, Peirani2008}, \cite{DelPopolo2021}, by \cite{Peirani2006, Peirani2008}. As done in Paper I, we will fit these data with the $J\eta \rm LT_{-2/3}$, (Eq. \ref{eq:Om}), and $J\eta \rm LT_{-1/3}$ (Eq. (\ref{eq:Omi})). 
%
%

\begin{figure*}[!ht] 
\includegraphics[scale=0.4]{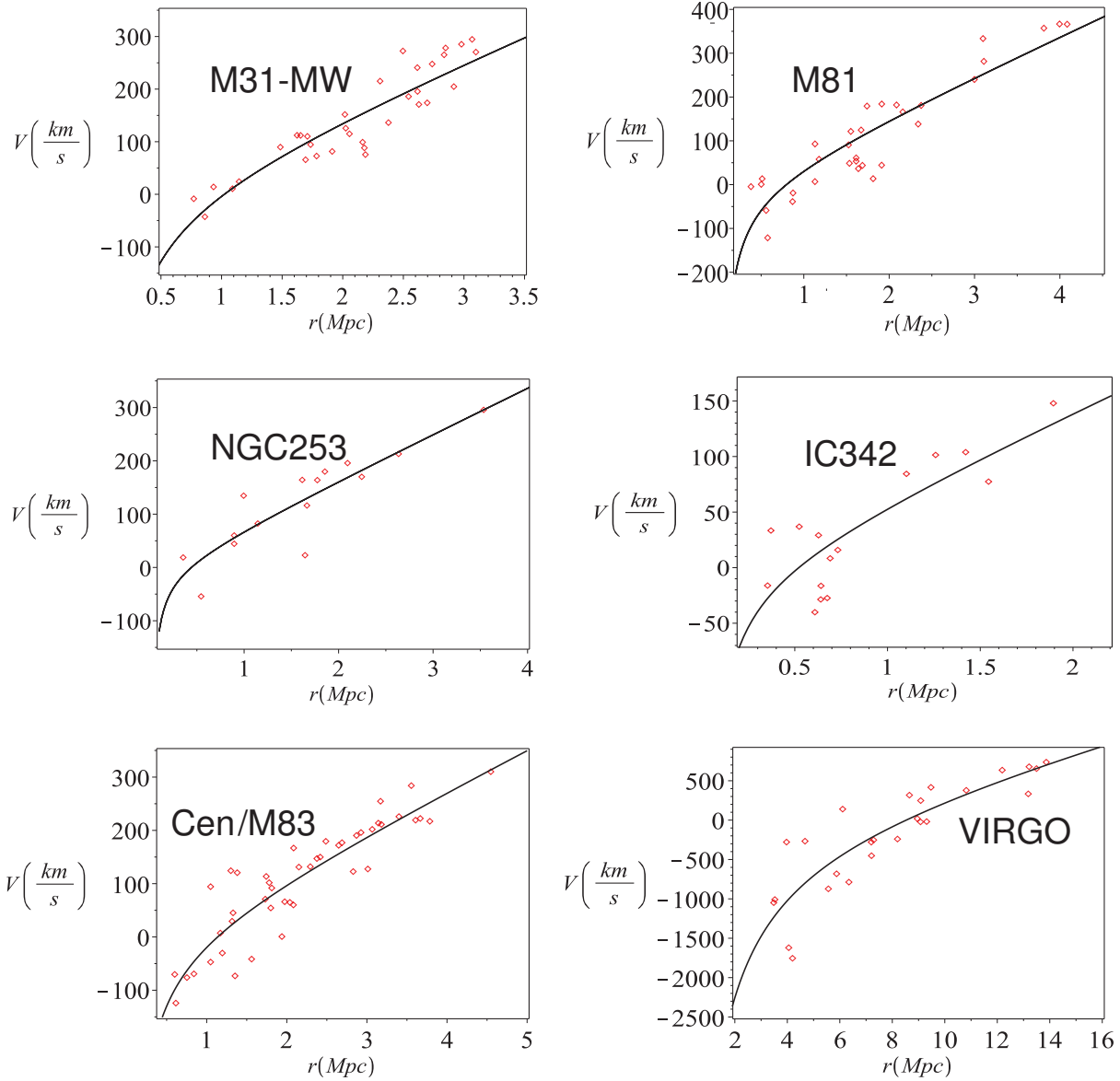}
	\caption{
		Velocity-distance plots for the groups of galaxies studied. The top left panel represents the M31-MW group, the top right panel the M81 group, the central left panel the NGC 253 group, the central right panel the IC 342 group, the bottom left panel the Cen/M83 group, and the bottom right panel the Virgo cluster. The red diamonds represent the data from \cite{Peirani2006,Peirani2008}. The black line is the fit to the data obtained with the model $J\eta \rm LT_{-2/3}$. 		
	}
	\label{fig:fig3}
\end{figure*}

Fig.\ref{fig:fig3} plots the $v-R$ relationships for the groups studied. From top to bottom we have that the top left panel represents the M31-MW group, the top right panel the M81 group, the central left panel the NGC 253 group, the central right panel the IC 342 group, the bottom left panel the Cen/M83 group, and the bottom right panel the Virgo cluster. The red diamonds represent the data from \cite{Peirani2006,Peirani2008}. The black line is the fit to the data. We represented only the case $w=-2/3$, because the other cases ($w=-1$, and $w=-1/3$), are almost indistinguishable from the case $w=-2/3$ with the dimension of the plots used. At small values of the radius the ordering of the plots is the same of that in Fig. 4:  $w=-1/3$, is upper than $w=-2/3$, and this last upper than $w=-1$.

\subsection{M31-MW}

The \cite{Karachentsev2002}
data were fitted with Eq. (\ref{eq:Om}) (case $J\eta \rm LT_{-2/3}$), and Eq. (\ref{eq:Omi}) (case $J\eta \rm LT_{-1/3}$). 
The results are shown in Table \ref{tab:hRM_fits}. By means of the SLT model \cite{Karachentsev2002} found a mass $1.5 \times 10^{12} M_{\odot}$, and   
a turn-around radius of $0.94 \pm 0.10$ Mpc. \cite{Peirani2006}, by means of the MLT model found $ (2.5 \pm 0.7) \times 10^{12} M_{\odot}$, $R_0=1.0 \pm 0.1$ Mpc, and $h=0.74 \pm 0.04$. The value of the mass of \cite{Peirani2006} is larger than that of  \cite{Karachentsev2002}, that used the SLT model. As already discussed in Paper I, the class of the LT models give a higher masses, and smaller $h$ if the effect of the cosmological constant, angular momentum, and other effects which contribute with positive terms in the equation of motion are taken into account. 
%
%
The values of $R_0$, in our cases ($J\eta \rm LT_{-2/3}$, $J\eta \rm LT_{-1/3}$) 
are in agreement, within the estimated uncertainties, with estimate reported in \cite{Peirani2006}.
Concerning the values of $h$, it is in agreement to that of \cite{Peirani2006} in all cases, while the mass $M$ obtained by \cite{Peirani2006} is slightly smaller. 
The errors reported, come from the fitting procedure.

\subsection{The M81 group}
The M81 group has been studied by many authors. For instance  \cite{Karachentsev2002},
\cite{Karachentsev2002a,Karachentsev2006} found $R_0 =0.89 \pm 0.05$ Mpc, and $M=(1.03 \pm 0.17) \times 10^{12} M_{\odot}$.
Their value of $R_0$ is in agreement to that of our models, while our mass is larger. 
\cite{Peirani2008} found $M=(0.92 \pm 0.24) \times 10^{12} M_{\odot}$, smaller than our cases and $h=0.67 \pm 0.04$, in agreement with our cases. 

\subsection{The NGC253 group}   
\cite{Karachentsev2003b} obtained $R_0=0.7 \pm 0.1$ Mpc, and $M=(5.5 \pm 2.2) \times 10^{11} M_{\odot}$, both larger than our estimates. \cite{Peirani2008} found $M=(1.3 \pm 1.8) \times 10^{11} M_{\odot}$, and $h=0.63 \pm 0.06$, both in agreement with our estimates. 

\subsection{The IC342 group}   
\cite{Karachentsev2003a} found that $R_0 =0.9 \pm 0.1$, and $M=(1.07 \pm 0.33) \times 10^{12} M_{\odot}$, both larger than our estimates. \cite{Peirani2008} found $M=(2.0 \pm 1.3) \times 10^{11} M_{\odot}$, $R_0$ ($\simeq 0.53$ Mpc), and $h=0.57 \pm 0.10$. Our values of mass, $R_0$, and $h$ agree with \cite{Peirani2008} estimates.      

\subsection{The CenA/M83 group}
\cite{Karachentsev2002b,Karachentsev2007} studied this group taking into account the effect of cosmological constant, and found $R_0=1.55 \pm 0.13$ Mpc, and $M=(6.4 \pm 1.8) \times 10^{12} M_{\odot}$, both larger than our estimates. \cite{Peirani2008}, found values 3-4 times smaller ($M=(2.1 \pm 0.5) \times 10^{12} M_{\odot}$), and $h=0.57 \pm 0.04$. Both the values of $M$, and $h$ that we found are in agreement with \cite{Peirani2008}. 

\subsection{The Virgo cluster}
Several authors studied this well known cluster with different methods. \cite{Hoffman1980}, \cite{Fouque_2001} used the SLT model, \cite{Tully1984} used the Virial theorem, obtaining masses smaller than $10^{15} M_{\odot}$. By converse, \cite{Fouque_2001} found a larger value ($1.3 \times 10^{15} M_{\odot}$). \cite{Peirani2006}, by means of the MLT model, 
got $M=(1.10 \pm 0.12) \times 10^{15} M_{\odot}$, smaller than our estimates,
$h=0.65 \pm 0.09$, and $R_0= 8.6 \pm 0.8$ Mpc, both in agreement with our estimates.
~\\
~\\

From the previous discussion, the estimates of mass, $M$, turn-around, $R_0$, and Hubble constant, are usually in agreement with those of \cite{Peirani2006,Peirani2008}. From $w=-1$ ($J\eta$LT case) to $w=-1/3$ ($J\eta \rm LT_{-1/3}$ case) the value of the Hubble constant, in units of 100 $\rm km s^{-1}/Mpc$, $h$, increases, and the mass, $M$, decreases (see Table \ref{tab:hRM_fits}). 
As discussed in Paper I, in the J$\eta$LT case ($w=-1$), some of the values of $h$ are smaller than the range of values in which the Hubble constant is constrained. From two decades ago to now the constraints on the Hubble constant has changed from  $72^{+8}_{-8}$ km/Mpc s, to the range $67-75$ km/Mpc s \citep{Freedman2019}. Other constraints from gravitational wave physics 
give $70.3^{+5.3}_{-5.0}$ km/Mpc s \citep{Hotokezaka2019}, and smaller range are obtained with $\rm DES+BAO+BBN$, $67.4^{+1.1}_{-1.2}$ km/Mpc s, and $67.5 \pm 1.1$ km/Mpc s \citep{Schoneberg2019}. The smaller value of all the previous constraints is $64$ km/Mpc s, which disagrees with the prediction of the J$\eta$LT for CenA/M83. The predictions of 
$J\eta \rm LT_{-2/3}$, and $J\eta \rm LT_{-1/3}$ disagree slightly with the ranges indicated, only in the case of CenA/M83. As discussed in Paper I, this issue could be due to non completeness of the data used in 2008 by \cite{Peirani2008}\footnote{This incompleteness is confirmed by the fact that a significant amount of faint dwarf galaxy candidates were discovered by means of a survey of the Centaurus group performed few years ago \citep{Muller2017}.}.

\begin{figure}[!ht] 
	\centering 
	\includegraphics[scale=0.35]{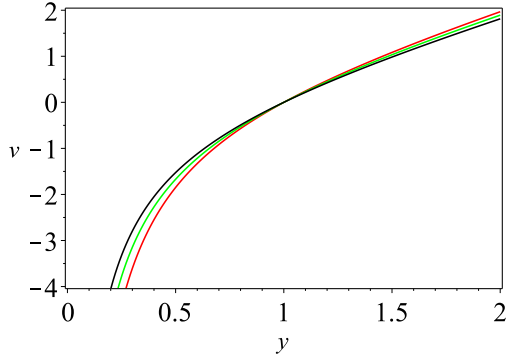}
	\caption{
		$v-R$ relationship for the $J\eta$LT model (red curve), $J\eta \rm LT_{-2/3}$ (green curve), and $J\eta \rm LT_{-1/3}$ (black curve). 
	}
	\label{fig:fig4}
\end{figure}



In Fig. (\ref{fig:fig4}), we plot the $v-R$ relations for the $J\eta$LT case (red line), the $J\eta \rm LT_{-2/3}$ case (green line), and the $J\eta \rm LT_{-1/3}$ case (black line). The plot shows that for distance smaller than $R_0$, the $J\eta$LT case has larger negative velocities with respect to $J\eta \rm LT_{-2/3}$ case, and this last larger negative velocities with respect to the $J\eta \rm LT_{-1/3}$ case. This means that turn-around happens before in the case $J\eta$LT, then in the $J\eta \rm LT_{-2/3}$ case, and finally in the $J\eta \rm LT_{-1/3}$ case.

\section{Comparison of the model with other results in literature} 
In this section, we want to compare the results of the model described in Section \ref{sec:Model}
with other results focusing on the effect of DE on structure formation. For example, \cite{Kuhlen2005} studied the effect of DE on the density profile of DM haloes, considering models with constant equation of state models. They found that larger values of $w$ 
give rise to higher concentrations and higher halo central densities. This is due to the fact that collapse happens earlier and because haloes have higher virial densities.  
Similarly, \cite{Klypin2003} studied the properties of dark matter haloes in a variety of DE models. They also found that DE halos are denser than those in $\Lambda$CDM because the DE haloes collapse earlier. This conclusion is in agreement with our model. As shown in Fig. 4, 
the turn-around and collapse happens earlier in larger $w$ cases. 

Here, in Fig. 5, we compare our results with the non-linear overdensity at collapse, 
$\Delta_{\rm vir}$, and the linear overdensity at collapse, $\delta_c$ in quintessence cosmologies, obtained by \cite{Kuhlen2005}.

\begin{figure}[!ht] 
	\centering 
	\includegraphics[scale=0.65]{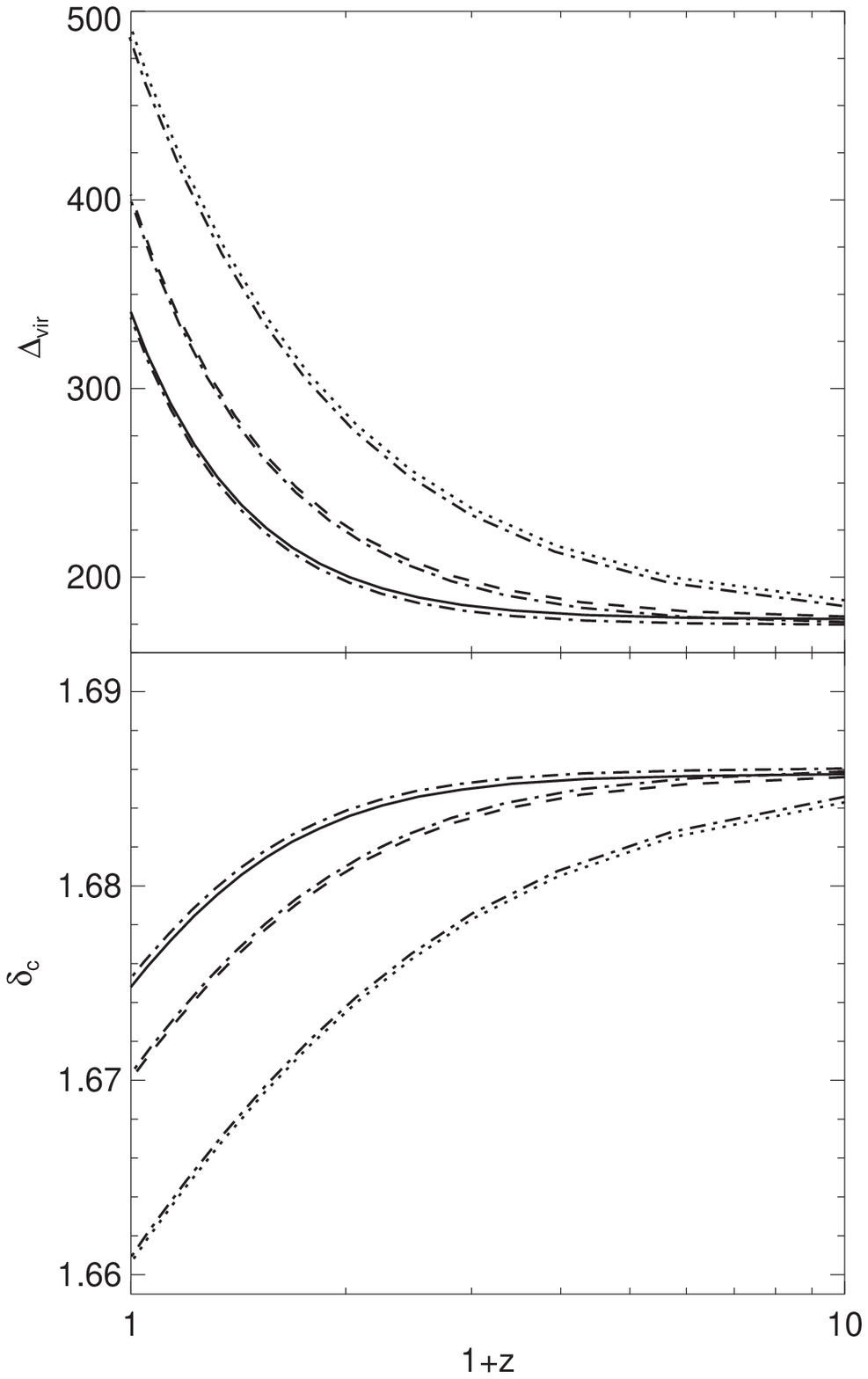}
	\caption{Plot of the linear and non-linear overdensities at collapse in quintessence cosmologies. 
Top panel: $\Delta_{\rm vir}$, the non-linear overdensity at collapse in terms of redshift $z$. 
The solid, dashed and dotted lines represent respectively the case $w=-1$ in the \cite{Kuhlen2005} paper, $w=-0.75$ in the \cite{Kuhlen2005} paper and $w=-0.50$ in the \cite{Kuhlen2005} paper, while the dot dashed lines represent the corresponding cases in our model. Bottom panel, 
$\delta_c$, the linear overdensity at collapse, as a function of redshift. The lines style is as that for the top panel.	}
\label{fig:fig4}
\end{figure}

In the top panel of Fig. 5, we plot the non-linear overdensity at collapse, $\Delta_{\rm vir}$, in terms of redshift $z$.
The solid, dashed and dotted lines represent respectively the case $w=-1$ in the \cite{Kuhlen2005} paper, $w=-0.75$ in the \cite{Kuhlen2005} paper and $w=-0.50$ in the \cite{Kuhlen2005} paper. In all cases, the dot dashed line represents that in our model. In the bottom panel, we plot the linear overdensity at collapse, $\delta_c$, as a function of redshift. The lines style is as that for the top panel.

As we told, the collapse happens earlier for larger values of $w$. The increase in $\Delta_{\rm vir}$ for larger $w$ results primarily from later collapse redshifts. The growth of structure is affected by the DE, more rapid for high values of $w$ (as shown in Fig. 4).  

\cite{Baushev2020}, studied the Hubble stream near a massive object, while 
\cite{Baushev2019}, gave an analytic solution for the relation between the mass of group and the turn-around radius $R_0$ taking into account of DE. He found that
\begin{equation}
M=2.278 \cdot 10^{12} \times (\frac{R_0}{\rm 1 Mpc})^3 (\frac{H_0}{\rm 73 km/s/Mpc})^2 
\label{eqq:kuhlen}
\end{equation}

In his papers he did not take account of shear and vorticity. As we showed in \cite{DelPopolo2020a}, the angular momentum can be expressed in terms of shear and vorticity (see Eq. 12). Then, in order to compare our result to that of \cite{Baushev2019} we neglect angular momentum and dynamical friction in our Eq. (\ref{eq:coll1}). Similarly, to compare our result to that of \cite{Kuhlen2005}, we used the parameters and the assumptions they used.

So, as discussed in \cite{DelPopolo2021}, and Eq. (\ref{eq:LTt}), in the case dynamical friction, shear and vorticity are neglected, the relation between mass and $R_0$ in Eq. (\ref{eq:LTt}), that can be written in terms of an Hubble constant equal to $73 \rm km/s/Mpc$ as
\begin{equation}
M=2.25 \times 10^{12} (\frac{R_0}{\rm 1 Mpc})^3 (\frac{H_0}{73 \rm km/s/Mpc})^2
\end{equation}
which is in agreement with Eq. (\ref{eqq:kuhlen}) of \cite{Kuhlen2005}.

\section{Constraints on the DM EoS parameter}
\label{sec:DMconstraints}

The turn-around, as already discussed, is the shell of the structure at which the velocity of expansion is zero. It is the region of space that delimits the region expanding with Hubble flow from that which will recollapse. Because of this particularity, several authors have proposed it as a promising way to test cosmological models \citep{Lopes2018}. It has been used to study dark energy models, to disentangle between those models, the $\Lambda$CDM model, and modified gravity models \citep{Pavlidou2014,Pavlidou2014a,Faraoni2015,Bhattacharya2017,Lopes2018,Lopes2019}.
 
%
%
In the majority of studies, the turn-around has been treated in the framework of General Relativity, or modified theories of gravity, as a geometric quantity. In real structures, the structure, and dimension of $R_0$ depends from several quantities like angular momentum, dynamical friction, or dark energy. It is then necessary to use a more physical model than those published. 
 
$R_0$ is also modified by the presence of vorticity, and shear in the equation of motion. 
In order to see how $R_0$ is modified, in \cite{DelPopolo2020}, we used an extended spherical collapse model (ESCM)  \citep{DelPopolo2013,DelPopolo2013a,Pace2014,Mehrabi2017,Pace2019}. 
In \cite{DelPopolo2020a}, we showed how taking also account of dynamical friction changed $R_0$.

In \cite{DelPopolo2020}, \cite{DelPopolo2020a}, we also showed how is possible to put some constraints on the DE EoS parameter $w$, by means of the $M-R_0$ plane. 

\begin{figure*}[!ht]
	\centering
	\includegraphics[width=12cm,angle=0]{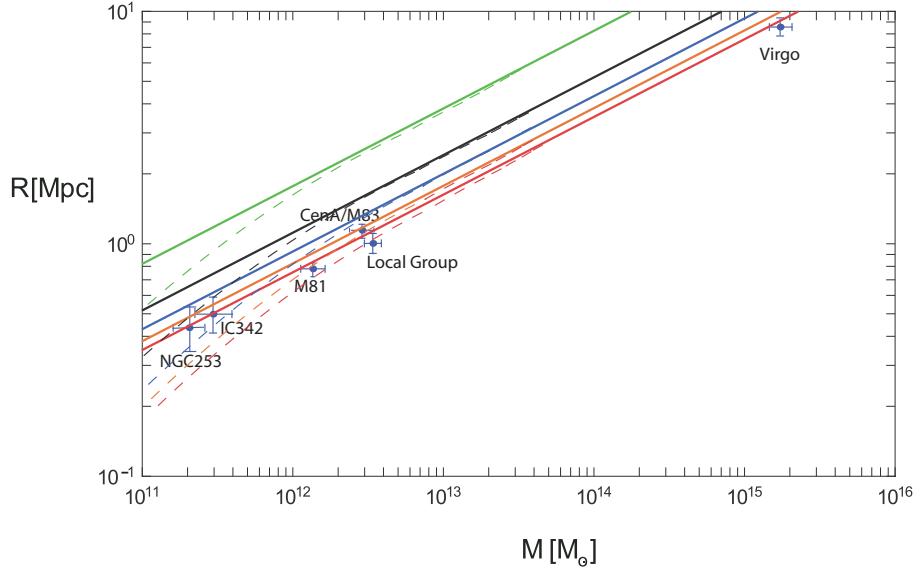}
	\caption{
Mass-radius relation of stable structures for different $w$. The solid lines, obtained by \cite{Pavlidou2014a}, from bottom to top
represent $w=-2.5$ (red solid line), -2 (pink solid line), -1.5 (blue solid line), -1 (black solid line), and -0.5 (solid green line).
The dashed lines correspond to the solid lines, but they were obtained taking the effect of the cosmological constant, angular momentum, and dynamical friction 
 \citep{DelPopolo2020a}. The dots with error bars come from the case $J\eta \rm LT_{-1/3}$, and are taken from Table~\ref{tab:hRM_fits}.
	}
	\label{fig:comparison}
\end{figure*}

~\\

The constraints on $w$ depends on the estimated values of the mass and $R_0$ of galaxies, groups, and clusters. 

In \cite{DelPopolo2020}, \cite{DelPopolo2020a}, we calculated the mass, and turn-around for several groups, and also studied how the $M-R_0$ relation is modified by shear, rotation, and dynamical friction. 

Here, we recalculate the constraints shown in \cite{DelPopolo2020}, \cite{DelPopolo2020a} by means of the revised value of mass, $M$, and $R_0$ presented in this paper.

Fig.\ref{fig:comparison} plots the mass-radius relation, $M-R_0$, of stable structures for different $w$. The solid lines from bottom to top represent $w=-2.5$ (red solid line), -2 (pink solid line), -1.5 (blue solid line), -1 (black solid line), and -0.5 (solid green line) (see \cite{Pavlidou2014a}). The dashed lines are the $M-R_0$ lines for the same values of $w$ described 
obtained using the model in \cite{DelPopolo2020a} considering the effect of shear, vorticity, angular momentum, dynamical friction, and dark energy only in the case $w=-1$. The dots with error bars, are data obtained in the previous sections, and reported in Table~\ref{tab:hRM_fits} (case J$\eta \rm LT_{-1/3}$).

\begin{table*}[t]
	\centering
	\caption{The allowed ranges of $w$.}
	\label{tab:table3}
	\begin{tabular}{@{}ll}
		\hline
		Stable structure  & range of $w$  \\
		\hline
		M81     &$w \ge -2$\\
		IC342   &$w \ge -1$\\
		NGC253  &$w \ge -0.9$\\
		CenA/M83    &$w \ge -1.6$\\
		Local Group &$w \ge -2.2$\\
		Virgo       &$w \ge -2.5$\\
		\hline
	\end{tabular}
\end{table*}

Table~\ref{tab:table3} reproduces the constraints to $w$. They differ from previous constraints as those obtained by \cite{Pavlidou2014}, \cite{Pavlidou2014a}.

In order to explain the difference between the results in \cite{Pavlidou2014}, \cite{Pavlidou2014a}, or \cite{DelPopolo2020a}, we recall the models used in those papers. As already discussed, in the majority of studies found in literature the turn-around is obtained starting from a metric in the framework of General Relativity, or modified theories of gravity. \cite{Pavlidou2014a} started from the metric in their Eq. 3.1. They derived the equation of evolution of the Universe. They wrote the equations in the case of a spherically symmetric configuration taking into account dark energy. They then determined the region in which matter can decouple from expansion and collapse. 
\cite{Pavlidou2014} followed a similar path. Turn-around is obtained using geometry, excluding  fundamental effects like tidal interaction, random angular momentum, dynamical friction, that are related to the content in baryons of the system. Real structures, like clusters of galaxies, are much more complicated than what the methods of \cite{Pavlidou2014a}, and \cite{Pavlidou2014} can describe. 
It is then necessary to use a more physical model than those published. 
This is the reason why in \cite{DelPopolo2020}, \cite{DelPopolo2020a} we used an extended spherical collapse model. The model takes into account shear, vorticity, dynamical friction, and DE. The effect of shear, vorticity, and dynamical friction, not taken into account in \cite{Pavlidou2014a}, and \cite{Pavlidou2014}, and also the majority of models dealing with the turn-around, change the relation between $M$, and $R_0$. 
Consequently the constraints obtained in \cite{DelPopolo2020}, \cite{DelPopolo2020a} differ from those in \cite{Pavlidou2014a}, and \cite{Pavlidou2014}.

%
%
~\\

A last observation, is that there is no correlation between the groups masses, and $w$. 
If one looks at NGC253, and IC342 which are the less massive, one finds values of $w= -0.9$, and -1, while in the case of Virgo, the most massive one gets -2.5, and one can think to a correlation. In reality a simple plot shows that there is not a strict correlation between mass and $w$.

\section{Conclusions}
\label{sec:Conclusions}

In this paper, we extended one of the models studied in Paper I, the $J\eta$LT model,  
to consider the effect of changing the parameter $w$ of the EoS of dark energy. 
The $J\eta$LT model, by converse, had been an extension of the SLT model taking into account cosmological constant, angular momentum, and dynamical friction. We showed that the change of $w$ modifies the evolution of perturbations. For example, turn-around happens before for more negative values of $w$. By solving the equation of motion, we got a relation between the mass, $M$, and the turn-around radius $R_0$, similarly to what done in \cite{Peirani2006,Peirani2008},
\cite{DelPopolo2021}. For a given, $R_0$, the perturbation mass of the case with $w=-1$ ($J\eta$LT) is larger than those having $w=-2/3$ ($J\eta \rm LT_{-2/3}$ case), and $w=-1/3$ ($J\eta \rm LT_{-1/3}$ case).
After finding a relation between mass, $M$, and turn-around, $R_0$, we found a velocity, $v$, radius, $R$, relation, depending on mass and the Hubble constant. The data of the local group, M81, NGC 253, IC342, CenA/M83, and Virgo were fitted with this relation obtaining best fitting values for the mass, $M$, and Hubble constant of the group considered. The mass decreases for less negative value of $w$, while the Hubble constant has the opposite behavior.  
Finally, constraints to $w$ were obtained from the mass, $M$, and turn-around radius obtained by the studied groups of galaxies, comparing with the $M-R_0$ relation obtained by \cite{DelPopolo2020a}. The constraints differ from previous ones \citep{Pavlidou2014,Pavlidou2014a}, because those predictions were 
based on the calculation of the mass, $M$, and $R_0$ by means of the virial theorem, and because the $M-R_0$ relationship did not account the effect of shear, rotation, and dynamical friction.

\begin{acknowledgments}
The authors gratefully thanks to S.Peirani and A. De Freitas Pacheco for fruitful discussion.
We also thank the anonymous referee for his suggestions that helped us to improve the paper.
\end{acknowledgments}

\bibliography{old_MasterBib}{}
\bibliographystyle{aasjournal}



\end{document}